\documentclass[aps,pra,showpacs,twocolumn,superscriptaddress,floatfix,fleqn,byrevtex]{revtex4}

\usepackage{amsmath,amssymb}
\usepackage{graphics,color,epsfig}

\begin{document}
\title{Negative-Index Metamaterials: Second-Harmonic Generation, Manley-Rowe
Relations and Parametric Amplification }
\author{A.~K.~Popov}\email{apopov@uwsp.edu}
\homepage{http://www.kirensky.ru/popov}\affiliation{Department of
Physics \& Astronomy and Department of Chemistry, University of
Wisconsin-Stevens Point, Stevens Point, WI 54481-3897}

\author{Vladimir M. Shalaev}\email{shalaev@purdue.edu}
\homepage{http://www.ece.purdue.edu/~shalaev} \affiliation{School of
Electrical and Computer Engineering, Purdue University,
West Lafayette, IN 47907-2035.}

\date{January 8, 2006}
\begin{abstract}\sloppy
Second harmonic generation and optical parametric amplification in
negative-index metamaterials (NIMs) are studied. The opposite
directions of the wave vector and the Poynting vector in NIMs results
in a "backward" phase-matching condition, causing significant changes
in the Manley-Rowe relations and spatial distributions of the coupled
field intensities. It is shown that absorption in NIMs can be
compensated by backward optical parametric amplification. The
possibility of distributed-feedback parametric oscillation with no
cavity has been demonstrated. The feasibility of the generation of
entangled pairs of left- and right-handed counter-propagating photons
is discussed.
\end{abstract}
 \maketitle

\section{Introduction}

\label{i} Recent demonstration of a negative refractive index for
metamaterials in the optical range \cite{NIMExp1,NIMExp2} opens new avenues
for optics and especially nonlinear optics. In parallel with progress for
metal-dielectric metamaterials, experimental demonstrations of negative
refraction in the near IR range have been made in a GaAs-based photonic
crystals \cite{pc1} and in Si-Polyimide photonic crystals \cite{pc2}.
Negative refractive-index metamaterials (NIMs) are also referred to as
left-handed materials (LHMs). The sufficient (but not necessary) condition
for a negative refractive index is simultaneously negative dielectric
permittivity $\epsilon (\omega )$ and negative magnetic permeability $\mu
(\omega )$ \cite{Vesel}. Negative magnetic permeability in the optical range
has been demonstrated in \cite{mu1,mu2,mu3}. NIMs exhibit highly unusual
optical properties and promise a great variety of unprecedented applications.
Optical magnetization, which is normally ignored in linear and
nonlinear-optics of the ordinary, positive-index materials (PIMs) (i.e.,
right-handed materials, RHMs) plays a crucial role in NIMs.

\sloppy The main emphasis in the studies of NIMs has been placed so
far on linear optical effects. Recently it has been shown that NIMs
including structural elements with non-symmetric current-voltage
characteristics can possess a nonlinear magnetic response at optical
frequencies \cite{Lap} and thus combine unprecedented linear and
nonlinear electromagnetic properties. Important properties of second
harmonic generation (SHG) in NIMs in the constant-pump approximation
were discussed in \cite{Agr} for semi-infinite materials and in
\cite{Lens} for a slab of a finite thickness. The propagation of
microwave radiation in nonlinear transmission lines, which are the
one-dimensional analog of NIMs, was investigated in \cite{Kozyr}. The
possibility of exact phase-matching for waves with counter-propagating
energy-flows has been shown in \cite{KivSHG} for the case when the
fundamental wave falls in the negative-index frequency domain and the
SH wave lies in the positive-index domain. The possibility of the
existence of multistable nonlinear effects in SHG was also predicted
in \cite{KivSHG}.

As seen from our consideration below, the phase-matching of normal and
backward waves is inherent for nonlinear optics of NIMs. We note here that
the important advantages of interaction schemes involving counter-directed
Poynting vectors in the process of optical parametric amplification in
ordinary RHMs were discussed in early papers \cite{Har}. However, in RHMs
such schemes impose severe limitations on the frequencies of the coupled waves
because of the requirement that one of the waves has to be in the
far-infrared range.

Absorption is one of the biggest problems to be addressed for
the practical applications of NIMs. In \cite{Agr,Lens}, a transfer of the
near-field image into SH frequency domain, where absorption is typically
much less, was proposed as a means to overcome dissipative losses and thus
enable the superlens.

In this paper, we demonstrate unusual characteristics in the spatial
distribution of the energy exchange between the fundamental and
second-harmonic waves. Both semi-infinite and finite-length NIMs are
considered and compared with each other and with ordinary PIMs.
Our analysis is based on the solution to equations for the
coupled waves propagating in lossless NIMs beyond the constant-pump
approximation. The Manley-Rowe relations for NIMs are analyzed and they are
shown to be strikingly different from those in PIMs. We also propose a new
means of compensating losses in NIMs by employing optical parametric
amplification (OPA). This can be realized by using control electromagnetic
waves (with frequencies outside the negative-index domain), which provide
the loss-balancing OPA inside the negative-index frequency domain. We also
predict laser oscillations without a cavity for frequencies in the
negative-index domain and the possibility of the generation of entangled pairs of
counter-propagating right- and left-handed photons.

The paper is organized as follows. Section \ref{shg} discusses the unusual
spatial distribution of the field intensities for SHG in finite and
semi-infinite slabs of NIMs. The Manley-Rowe relations are derived and
discussed here. The feasibility of compensating losses in NIMs by using the
OPA is considered in Section \ref{opa}. In this Section we also study
cavity-less oscillations based on distributed feedback. Finally, a summary
of the obtained results concludes the paper.

\section{Second harmonic generation in NIMs}

\label{shg}

\subsection{Wave vectors and Poynting vectors in NIMs}

\label{pv} We consider a loss-free material, which is left-handed at the
fundamental frequency $\omega_1$ ($\epsilon_1<0$, $\mu_1<0$), whereas it is
right-handed at the SH frequency $\omega_2=2\omega_1$ ($\epsilon_2>0$, $%
\mu_2>0$).
The relations between the vectors of the electrical, $\mathbf{E}$, and magnetic, $%
\mathbf{H}$, field components and the wave-vector $\mathbf{k}$ for a
traveling electromagnetic wave,
\begin{eqnarray}
\mathbf{E}(\mathbf{r},t)&=&\mathbf{E}_0(\mathbf{r})\exp[-i(\omega t-\mathbf{%
k\cdot r})]+ c.c.,  \label{EM} \\
\mathbf{H}(\mathbf{r},t)&=&\mathbf{H}_0(\mathbf{r})\exp[-i(\omega t -\mathbf{%
k\cdot r})]+ c.c.,  \label{HM}
\end{eqnarray}
are given by the following formulas
\begin{eqnarray}
\mathbf{k}\times\mathbf{E}& = &({\omega}/{c})\mu\mathbf{H},\quad\mathbf{k}%
\times\mathbf{H} =- ({\omega}/{c}) \epsilon\mathbf{E},  \label{kh} \\
\sqrt{\epsilon}{E}(\mathbf{r},t)&=&-\sqrt{\mu}{H}(\mathbf{r},t),  \label{eh}
\end{eqnarray}
which follow from Maxwell's equations. These expressions show that
the vector triplet $\mathbf{E}$, $\mathbf{H}$ and $\mathbf{k}$ forms
a right-handed system for the SH wave and a left-handed system for
the fundamental beam. Simultaneously negative $\epsilon_i<0$ and
$\mu_i<0$ result in a negative refractive index $n= -
\sqrt{\mu\epsilon}$. As seen from Eqs. (\ref{EM}) and
(\ref{HM}), the phase velocity $\mathbf{v}_{ph}$ is co-directed with $%
\mathbf{k}$ and is given by $\mathbf{v}_{ph}=({\mathbf{k}}/{k})({\omega}/{%
k})=({\mathbf{k}}/{k})({c}/{|n|})$, where ${k}^{2}=n^{2}(\omega/{c})^{2}$.
In contrast, the direction of the energy flow (Poynting vector) $\mathbf{S}$
with respect to $\mathbf{k}$ depends on the signs of $\epsilon $ and $\mu $:
\begin{eqnarray}
\mathbf{S}(\mathbf{r},t) &=&\frac{c}{4\pi}[\mathbf{E}\times\mathbf{H}] =%
\frac{c^{2}}{4\pi\omega\epsilon}[\mathbf{H}\times\mathbf{k}\times\mathbf{H}]
=  \notag \\
&=&\frac{c^{2}\mathbf{k}}{4\pi\omega\epsilon}H^{2} =\frac{c^{2}\mathbf{k}}{%
4\pi\omega\mu}E^{2}.  \label{s}
\end{eqnarray}
As mentioned, we assume here that all indices of $\epsilon$, $\mu$
and $n$ are real numbers. Thus, the energy flow $\mathbf{S}_1$ at
$\omega_1$ is directed opposite to $\mathbf{k}_1$, whereas
$\mathbf{S}_2$ is co-directed with $\mathbf{k}_2$.

\subsection{SHG: Basic equations and the Manley-Rowe relations}

\label{be} 
\begin{figure}[h]
\includegraphics[height=.3\textwidth]{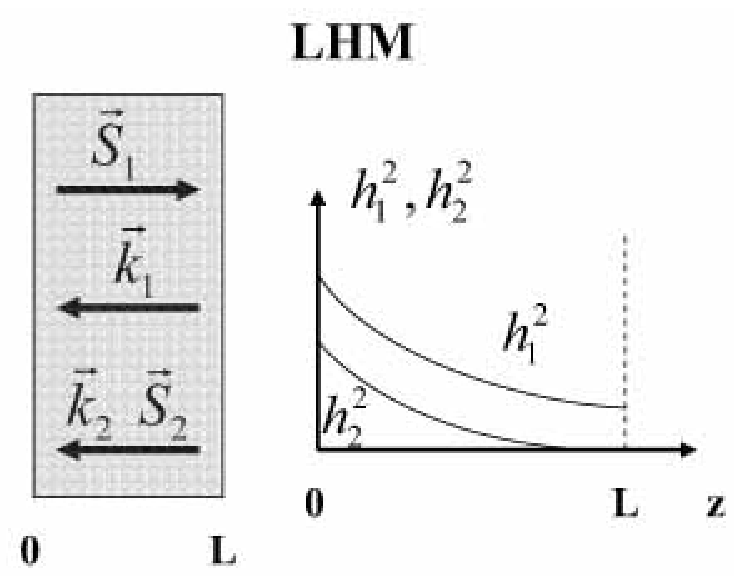}\newline
\includegraphics[height=.3\textwidth]{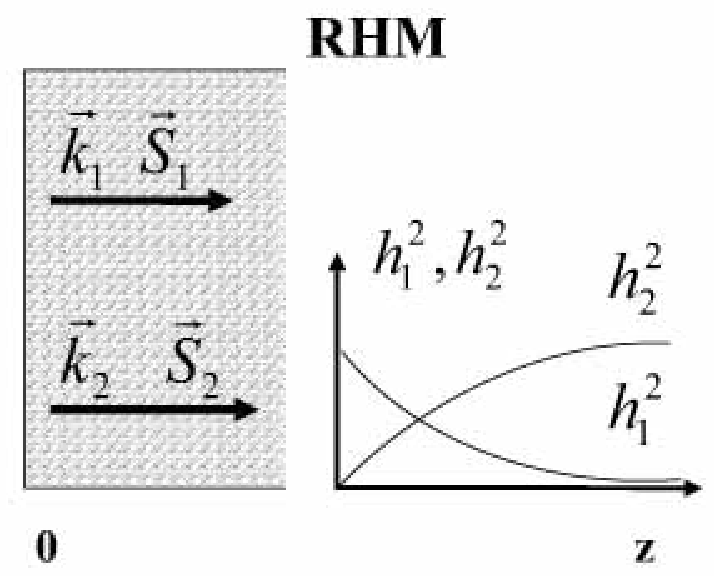}
\caption{SHG geometry and the difference between SHG in LHM and RHM slabs. }
\label{f1}
\end{figure}

We assume that an incident flow of fundamental radiation
$\mathbf{ S}_{1}$ at $\omega _{1}$ propagates along the z-axis,
which is normal to the surface of a metamaterial. According to
(\ref{s}), the phase of the wave at $\omega _{1}$ travels in the
reverse direction inside the NIM (the upper part of Fig.\ref{f1}).
Because of the phase-matching requirement, the generated SH
radiation also travels backward with energy flow in the same
backward direction. This is in contrast with the standard coupling
geometry in a PIM (the lower part of Fig.\ref{f1}).
Following the method of \cite{KivSHG}, we assume that a nonlinear
response is primarily associated with the magnetic component of the
waves. Then the equations for the coupled fields inside a NIM in the
approximation of slow-varying amplitudes acquire the form:
\begin{eqnarray}
\frac{dA_{2}}{dz} &=&i\dfrac{\epsilon _{2}\omega
_{2}^{2}}{k_{2}c^{2}}4\pi
\chi _{eff}^{(2)}A_{1}^{2}\exp (-\Delta kz),  \label{A2} \\
\frac{dA_{1}}{dz} &=&i\dfrac{\epsilon _{1}\omega
_{1}^{2}}{k_{1}c^{2}}8\pi \chi _{eff}^{(2)}A_{2}A_{1}^{\ast }\exp
(\Delta kz).  \label{A1}
\end{eqnarray}
Here, $\chi _{eff}^{(2)}$ is the effective nonlinear susceptibility,
$\Delta k=k_{2}-2k_{1}$ is the phase mismatch, and $A_{2}$ and
$A_{1}$ are the slowly varying amplitudes of the waves with the
phases traveling against the z-axis:
\begin{equation}
{H}_{j}(z,t)={A}_{j}\exp [-i(k_{j}z+\omega _{j}t)]+c.c.,  \label{Az}
\end{equation}
where, $\omega _{2}=2\omega _{1}$ and $k_{1,2}>0$ are the moduli of
the wave-vectors directed against the z-axis. We note that according
to Eq. (\ref{eh}) the corresponding equations for the electric
components can be written in a similar form, with $\epsilon _{j}$
substituted by $\mu _{j}$ and vice versa. The factors $\mu _{j}$
were usually assumed to be equal to one in similar equations for
PIMs. However, this assumption does not hold for the case of NIMs,
and this fact dramatically changes many conventional electromagnetic
relations. The Manley-Rowe relations \cite{MR} for the field
intensities and for the energy flows follow from Eqs. (\ref{s}) -
(\ref{A1}):
\begin{equation}
\frac{k_{1}}{\epsilon
_{1}}\dfrac{d|A_{1}|^{2}}{dz}+\frac{k_{2}}{2\epsilon
_{2}}\dfrac{d|A_{2}|^{2}}{dz}=0,\quad
\dfrac{d|S_{1}|^{2}}{dz}-\dfrac{d|S_{2}|^{2}}{dz}=0.
\label{MR1}
\end{equation}
The latter equation accounts for the difference in the signs of
$\epsilon _{1}$ and $\epsilon _{2}$, which brings radical changes to
the spatial dependence of the field intensities discussed below.

We focus on the basic features of the process and ignore the
dissipation of both waves inside the nonlinear medium; in addition,
we assume that the phase matching condition $k_{2}=2k_{1}$ is
fulfilled. The spatially-invariant form of the Manley-Rowe relations
follows from equation (\ref{MR1}):
\begin{equation}
|A_{1}|^{2}/\epsilon _{1}+|A_{2}|^{2}/\epsilon _{2}=C,  \label{I}
\end{equation}
where $C$ is an integration constant. With $\epsilon _{1}=-\epsilon
_{2}$, which is required for the phase matching, equation (\ref{I})
predicts that the \emph{difference} between the squared amplitudes
remains constant through the sample
\begin{equation}
|A_{1}|^{2}-|A_{2}|^{2}=C,  \label{D}
\end{equation}
as schematically depicted in the upper part of Fig. \ref{f1}. This
is in striking difference with the requirement that the \emph{sum}
of the squared amplitudes is constant in the analogous case in a
PIM, as schematically shown in the lower part of Fig. \ref{f1}. We
introduce now the real phases and amplitudes as $A_{1,2}=h_{1,2}\exp
(i\phi_{1,2})$. Then the equations for the phases, which follow from
Eqs. (\ref{A2}) and (\ref{A1}), show that if any of the fields
becomes zero at any point, the integral (\ref{I}) corresponds to the
solution with the constant phase difference $2\phi_{1}-\phi_{2}=\pi
/2$ over the entire sample.

The equations for the slowly-varying amplitudes corresponding to the
ordinary coupling scheme in a PIM, shown in the lower part of Fig.
\ref{f1}, are readily obtained from Eqs. (\ref{A2}) - (\ref{Az}) by
changing the signs of $k_{1}$ and $k_{2}$. This does not change the
integral (\ref{I}); more importantly, the relation between
$\epsilon_{1}$ and $\epsilon_{2}$ required by the phase matching now
changes to $\epsilon_{1}=\epsilon_{2}$, where both constants are
positive. The phase difference remains the same. Because of the
boundary conditions $h_{1}(0)=h_{10}$ and $h_{2}(0)=h_{20}=0$, the
integration constant becomes $C=h_{10}^{2}$. Thus, the equations for
the real amplitudes in the case of a PIM acquire the form:
\begin{eqnarray}
&&h_{1}(z)=\sqrt{h_{10}^{2}-h_{2}(z)^{2}},  \label{D3} \\
&&dh_{2}/{dz}=\kappa \lbrack h_{10}^{2}-h_{2}(z)^{2}],  \label{h24}
\end{eqnarray}
with the known solution
\begin{eqnarray}
h_{2}(z) &=&h_{10}\tanh (z/z_{0}),  \label{rhm2} \\
h_{1}(z) &=&h_{10}/\cosh (z/z_{0}),\,z_{0}=[\kappa h_{10}]^{-1}.
\label{rhm1}
\end{eqnarray}
Here, $\kappa =({\epsilon_{2}\omega_{2}^{2}}/{k_{2}c^{2}})4\pi
\chi_{eff}^{(2)}$. \emph{The solution has the same form for an
arbitrary slab thickness}, as shown schematically in the lower part of
Fig. \ref{f1}.

\subsection{SHG in a NIM slab}

\label{sl} 
\begin{figure}[h]
\includegraphics[width=.4\textwidth]{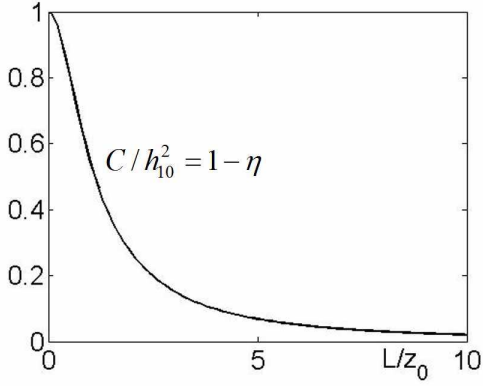}
\caption{The normalized integration constant $C/h_{10}^{2}$ and the energy
conversion efficiency $\protect\eta $ vs the normalized length of a NIM slab.}
\label{f2}
\end{figure}

\begin{figure}[h]
\includegraphics[height=.38\textwidth]{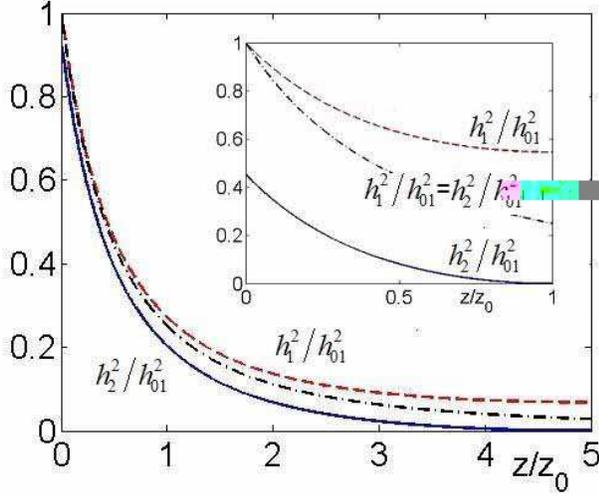}
\caption{The squared amplitudes for the fundamental wave (the dashed
line) and SHG (the solid line) in a lossless NIM slab of a finite
length. Inset: the slab has a length equal to one conversion length.
Main plot: the slab has a length equal to five conversion lengths.
The dash-dot lines show the energy-conversion for a semi-infinite
NIM.} \label{f3}
\end{figure}

Now consider phase-matched SHG in a lossless NIM slab of
a finite length L. Equations (\ref{A2}) and (\ref{D}) take the form:
\begin{eqnarray}
h_{1}(z)^{2} &=&C+h_{2}(z)^{2},  \label{D1} \\
dh_{2}/{dz} &=&-\kappa \lbrack C+h_{2}(z)^{2}].  \label{h12}
\end{eqnarray}
Taking into account the \emph{different boundary conditions in a NIM
as compared to a PIM}, $h_{1}(0)=h_{10}$ and $h_{2}(L)=0$, the
solution to these equations is as follows
\begin{eqnarray}
h_{2} &=&\sqrt{C}\tan [\sqrt{C}\kappa (L-z)],  \label{h22} \\
h_{1} &=&\sqrt{C}/\cos [\sqrt{C}\kappa (L-z)],  \label{h11}
\end{eqnarray}
where the integration parameter $C$ \emph{depends on the slab thickness $L$}
and on the amplitude of the incident fundamental radiation as
\begin{equation}
\sqrt{C}\kappa L=\cos ^{-1}(\sqrt{C}/h_{10}).  \label{C}
\end{equation}
Thus, \emph{the spatially invariant field intensity difference
between the fundamental and SH waves in NIMs depends on the slab
thickness, which is in strict contrast with the case in PIMs.} As
seen from equation (\ref{D1}), the integration parameter
$C=h_{1}(z)^{2}-h_{2}(z)^{2}$ now represents the deviation of the
conversion efficiency $\eta =h_{20}^{2}/h_{10}^{2}$ from unity:
$(C/h_{10}^{2})=1-\eta $. Figure \ref{f2} shows the dependence of
this parameter on the conversion length $z_{0}=(\kappa
h_{10})^{-1}$.
The figure shows that for the conversion length of 2.5, the NIM
slab, which acts as nonlinear mirror, provides about 80\% conversion
of the fundamental beam into a reflected SH wave. Figure \ref{f3}
depicts the field distribution along the slab. One can see from the
figure that with an increase in slab length (or intensity of the
fundamental wave), the gap between the two plots decreases while the
conversion efficiency increases (comparing the main plot and the
inset).

\subsection{SHG in a semi-infinite NIM}

\label{si} Now we consider the case of a semi-infinite NIM at $z>0$.
Since both waves disappear at $z\rightarrow \infty $ due to the
entire conversion of the fundamental beam into SH, $C=0$. Then
equations (\ref{D1}) and (\ref {h12}) for the amplitudes take the
simple form
\begin{eqnarray}
&&h_{2}(z)=h_{1}(z),  \label{D2} \\
&&dh_{2}/{dz}=-\kappa h_{2}^{2}.  \label{h23}
\end{eqnarray}
Equation (\ref{D2}) indicates 100\% conversion of the incident
fundamental wave into the reflected second harmonic at $z=0$ in a
lossless semi-infinite medium provided that the phase matching
condition $\Delta k=0$ is fulfilled. The integration of (\ref{h23})
with the boundary condition $h_{1}(0)=h_{10}$ yields
\begin{equation}
h_{2}(z)=\dfrac{h_{10}}{(z/z_{0})+1},\,z_{0}=(\kappa h_{10})^{-1}.
\label{si}
\end{equation}
Equation (\ref{si}) describes a \emph{concurrent decrease of both
waves of equal amplitudes along the z-axis;} this is shown by the
dash-dot plots in Fig. \ref{f3}. For $z\gg z_{0}$, the dependence is
inversely proportional to $z$. These \emph{spatial dependencies,
shown in Fig. \ref{f3}, are in strict contrast with those for the
conventional process of SHG in a PIM}, which are known from various
textbooks (compare, for example, with the lower part of
Fig.\ref{f1}).

\section{Optical parametric amplification and difference-frequency
generation in a NIM slab with absorption}

\label{opa}

\subsection{OPA: basic equations and Manley-Rowe relations}

\label{mrr} 
\begin{figure}[h]
\includegraphics[height=.3\textwidth]{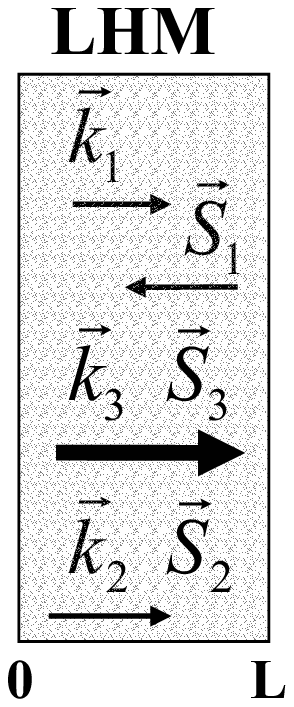} \includegraphics[height=.3%
\textwidth]{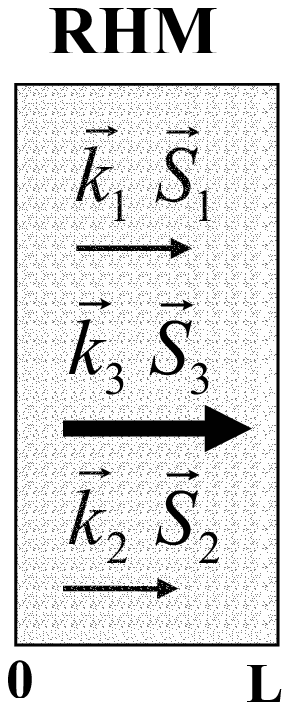}
\caption{The difference between OPA processes in LHM and PIM
slabs.}
\label{fig1}
\end{figure}

As mentioned in Subsection \ref{pv}, $\mathbf{S}$ is
counterdirected with respect to $\mathbf{k}$ in NIMs, because
$\epsilon <0$ and $\mu <0$. We assume that a left-handed wave at
$\omega _{1}$ travels with its wave-vector directed along the
$z$-axis. Then its energy flow $\mathbf{S}_{1}$ is directed against
the $z$-axis. We also assume that the sample is illuminated by a
higher-frequency electromagnetic wave traveling along the axis $z$.
The frequency of this radiation $\omega_{3}$ falls in a positive
index range. The two coupled waves with co-directed wave-vectors
$\mathbf{k}_{3}$ and $\mathbf{k}_{1}$ generate a
difference-frequency idler at $\omega_{2}=\omega_{3}-\omega_{1}$,
which has a positive refractive index. The idle wave contributes
back into the wave at $\omega_{1}$ through three-wave coupling and
thus enables optical parametric amplification (OPA) at $\omega_{1}$
by converting the energy of the pump field at $\omega_{3}$. Thus,
the nonlinear-optical process under consideration involves
three-wave mixing with wave-vectors co-directed along $z$. Note that
the energy flow of the signal wave, $\mathbf{S}_{1}$, is directed
against $z$, i.e., it is directed opposite to the energy flows of
the two other waves, $\mathbf{S}_{2}$ and $\mathbf{S}_{3}$ (Fig.
\ref{fig1}, the left part). Such a coupling scheme is in contrast
with the ordinary phase-matching scheme for OPA, which is
schematically shown in the right part of Fig. \ref {fig1}.
As above, we consider the magnetic type of the quadratic
nonlinearity. For the magnetic field
\begin{equation}
{H}_{j}(z,t)={h}_{j}\exp [i(k_{j}z-\omega _{j}t)]+c.c.,  \label{hz1}
\end{equation}
the nonlinear magnetization at the signal and idler frequencies is given by
the equations
\begin{eqnarray}
M_{1}^{NL} &=&2\chi_{eff}^{(2)}h_{3}h_{2}^{\ast }\exp
\{i[(k_{3}-k_{2})-\omega_{1}t]\},  \label{m1n1} \\
M_{2}^{NL} &=&2\chi_{eff}^{(2)}h_{3}h_{1}^{\ast }\exp
\{i[(k_{3}-k_{1})-\omega_{2}t]\}.  \label{m2n1}
\end{eqnarray}
Here, $j=1,2,3;\, \omega_{2}=\omega_{3}-\omega_{1};\,$ and
$k_{j}=|n_{j}|\omega _{j}/c>0$. Then the equations for the
slowly-varying amplitudes of the signal and the idler acquire the
form
\begin{eqnarray}
\frac{dh_{1}}{dz} &=&i\sigma_{1}h_{3}h_{2}^{\ast }\exp [i\Delta kz]+\frac{%
\alpha _{1}}{2}h_{1},  \label{h11} \\
\frac{dh_{2}}{dz} &=&i\sigma_{2}h_{3}h_{1}^{\ast }\exp [i\Delta kz]-\frac{%
\alpha_{2}}{2}h_{2},  \label{h21}
\end{eqnarray}
where $\sigma_{j}=8\pi
\chi_{eff}^{(2)}{\epsilon_{j}\omega_{j}^{2}}/{k_{j}c^{2}}$, $\Delta
k=k_{3}-k_{2}-k_{1}$, and $\alpha _{j}$ are the absorption indices.
The amplitude of the pump $h_{3}$ is assumed constant. We note the
following \emph{three fundamental differences} in equation (\ref
{h11}) as compared with the ordinary difference-frequency generation
(DFG) through the three-wave mixing of co-propagating waves in a PIM.
First, the sign of $\sigma_{1}$ is opposite to that of $\sigma_{2}$
because $ \epsilon_{1}<0$. Second, the opposite sign appears with
$\alpha_{1}$ because the energy flow $\mathbf{S_{1}}$ is directed
against the $z$-axis. Third, the boundary conditions for $h_{1}$ are
defined at the opposite side of the sample as compared to $h_{2}$ and
$h_{3}$ because their energy-flows $\mathbf{S_{1}}$ and
$\mathbf{S_{2}}$ are counter-directed.

At $\alpha_{1}=\alpha_{2}=0$, one finds with the aid of Eqs.
(\ref{h11}), (\ref{h21}) and (\ref{s}):
\begin{eqnarray}
&&\frac{d}{dz}\left[ \dfrac{S_{1z}}{\hbar {\omega_{1}}}-\dfrac{S_{2z}}{%
\hbar {\omega_{2}}}\right] =0,  \label{MR1} \\
&&\dfrac{d}{dz}\left[\sqrt{\dfrac{\mu_{1}}{\epsilon_{1}}}\dfrac{
|h_{1}|^{2}}{\omega_{1}}+\sqrt{\dfrac{\mu_{2}}{\epsilon
_{2}}}\dfrac{|h_{2}|^{2}}{\omega_{2}}\right] =0.  \label{MR21}
\end{eqnarray}
These equations represent the Manley-Rowe relations \cite{MR}, which
describe the creation of pairs of \emph{entangled counter-propagating
photons} $\hbar{\omega_{1}}$ and $\hbar{\omega_{2}}$. The equations
account for the opposite sign of the corresponding derivatives with
respect to z. Equation (\ref{MR21}) predicts that the \emph{sum} of
the terms proportional to the squared amplitudes of signal and idler
remains constant through the sample, which is in contrast with the
requirement that the \emph{difference} of such terms is constant in
the analogous case in a PIM. We note that according to Eqs. (\ref{eh})
and (\ref{s}) the corresponding equations for the electric components
in the case of the quadratic electric nonlinearity can be written in a
similar form with $\epsilon_{j}$ substituted by $\mu_{j}$. As seen
from the equations below, this does not change either the results
obtained or the main conclusions presented here; the same is true for
the case of SHG. As mentioned in Section \ref{shg}, the factors
$\mu_{j}$ were usually assumed equal to unity in equations for PIMs,
which is not the case for NIMs.

\subsection{OPA and DFG in NIMs}

\sloppy We introduce the normalized amplitudes
$a_{j}=\sqrt[4]{{\epsilon_{j}}/{\mu _{j}}}{h_{j}}/\sqrt{\omega _{j}};$
their squared values are proportional to the number of photons at the
corresponding frequencies. The corresponding equations for such
amplitudes acquire the form
\begin{eqnarray}
\dfrac{da_{1}}{dz} &=&-i{g}a_{2}^{\ast }\exp [i\Delta kz]+\frac{\alpha_{1}}{%
2}a_{1},  \label{a11} \\
\dfrac{da_{2}}{dz} &=&i{g}a_{1}^{\ast }\exp [i\Delta kz]-\frac{\alpha_{2}}{2%
}a_{2},  \label{a21}
\end{eqnarray}
where
${g}=(\sqrt{\omega_1\omega_2}/\sqrt[4]{\epsilon_1\epsilon_2/\mu_1\mu_2})
({8\pi}/{c}){\chi^{(2)}}h_{3}$. Accounting for the boundary conditions
$a_{1}(z=L)=a_{1L}$, and $a_{2}(z=0)=a_{20}$ (where $L$ is the slab
thickness), the solutions to equations (\ref{a11}) and (\ref{a21}) are
as follows
\begin{eqnarray}
a_{1}(z) &=&A_{1}\exp [(\beta_{1}+i\frac{\Delta k}{2})z]+  \notag \\
&+&A_{2}\exp [(\beta_{2}+i\frac{\Delta k}{2})z],  \label{a1z1} \\
a_{2}^{\ast }(z) &=&\kappa_{1}A_{1}\exp [(\beta_{1}-i\frac{\Delta
k}{2})z]+
\notag \\
&+&\kappa_{2}A_{2}\exp [(\beta_{2}-i\frac{\Delta k}{2})z].
\label{a2z1}
\end{eqnarray}
Here,
\begin{eqnarray}
&&\beta_{1,2}={(\alpha_{1}-\alpha_{2})}/{4}\pm
iR,\,\kappa_{1,2}=[\pm {R}+is]/g,  \label{bet1} \\
&&R=\sqrt{g^{2}-s^{2}},\,s=({\alpha_{1}+\alpha_{2}})/{4}-i{\Delta
k}/{2},\label{r1} \\
&&A_{1}=\{a_{1L}\kappa_{2}-a_{20}^{\ast}\exp
[(\beta_{2}+i\frac{\Delta k}{2})L]\}/D, \label{A11} \\
&&A_{2}=-\{a_{1L}\kappa_{1}-a_{20}^{\ast}\exp
[(\beta_{1}+i\frac{\Delta k}{2})L]\}/D,  \label{A21} \\
&&D=\kappa_{2}\exp [(\beta_{1}+i\frac{\Delta k}{2})L]-\kappa_{1}\exp
[(\beta_{2}+i\frac{\Delta k}{2})L].\,\qquad \label{D1}
\end{eqnarray}

At $a_{20}=0$, the \emph{amplification factor for the left-handed
wave} is given by $\eta_{a}(\omega_{1})=\left\vert
{a_{10}}/{a_{1L}}\right\vert ^{2}$, where
\begin{equation}
\frac{a_{10}}{a_{1L}}=\dfrac{\exp \left[ -\left(
\dfrac{\alpha_{1}-\alpha_{2}}{4}+i\dfrac{\Delta k}{2}\right)
L\right]}{\cos
RL+\left[\dfrac{\alpha_{1}+\alpha_{2}}{4R}-i\dfrac{\Delta
k}{2R}\right] \sin RL}.  \label{amp11}
\end{equation}
Alternatively, at $a_{1L}^{\ast }=0$, the \emph{conversion factor
for the difference-frequency generation of the left-handed wave} is
found as $\eta_{g}(\omega_{1})=\left\vert {a_{10}}/{a_{20}^{\ast
}}\right\vert ^{2}$, where
\begin{equation}
\frac{a_{10}}{a_{20}^{\ast }}=\dfrac{-(g/R)\sin RL}{\cos RL+\left[
\dfrac{\alpha_{1}+\alpha_{2}}{4R}-i\dfrac{\Delta k}{2R}\right] \sin
RL}. \label{gen11}
\end{equation}
Equation (\ref{amp11}) shows that the amplification of the
left-handed wave can be turned into a \emph{cavity-less oscillation}
when the denominator tends to zero. The conversion factor for DFG,
$\eta_{g}$, experiences a similar increase. In the case of $\Delta
k=0$ and small optical losses $(\alpha_{1}+\alpha_{2})L\ll \pi$,
equations (\ref{a1z1}) and (\ref{a2z1}) are reduced to
\begin{eqnarray}
a_{1}^{\ast }(z) &\approx &\frac{a_{1L}^{\ast }}{\cos (gL)}\cos
({gz})+\frac{
ia_{20}}{\cos (gL)}\sin [{g(z-L)}],\quad   \label{lla11} \\
a_{2}(z) &\approx &\frac{ia_{1L}^{\ast }}{\cos (gL)}\sin
({gz})+\frac{a_{20} }{\cos (gL)}\cos [{g(z-L)}].\quad
\label{lgen21}
\end{eqnarray}
The output amplitudes are then given by
\begin{eqnarray}
a_{10}^{\ast } &=&\frac{a_{1L}^{\ast }}{\cos (gL)}-ia_{20}\tan ({gL}),
\label{a101} \\
a_{2L} &=&ia_{1L}^{\ast }\tan ({gL})+\frac{a_{20}}{\cos ({gL})}.
\label{a2l1}
\end{eqnarray}
Thus, the oscillation threshold value for the control field intensity in
this case is given by $g_{t}=\pi /2L$. It increases with absorption and
phase mismatch.
\begin{figure}[h!]
\includegraphics[width=.35\textwidth]{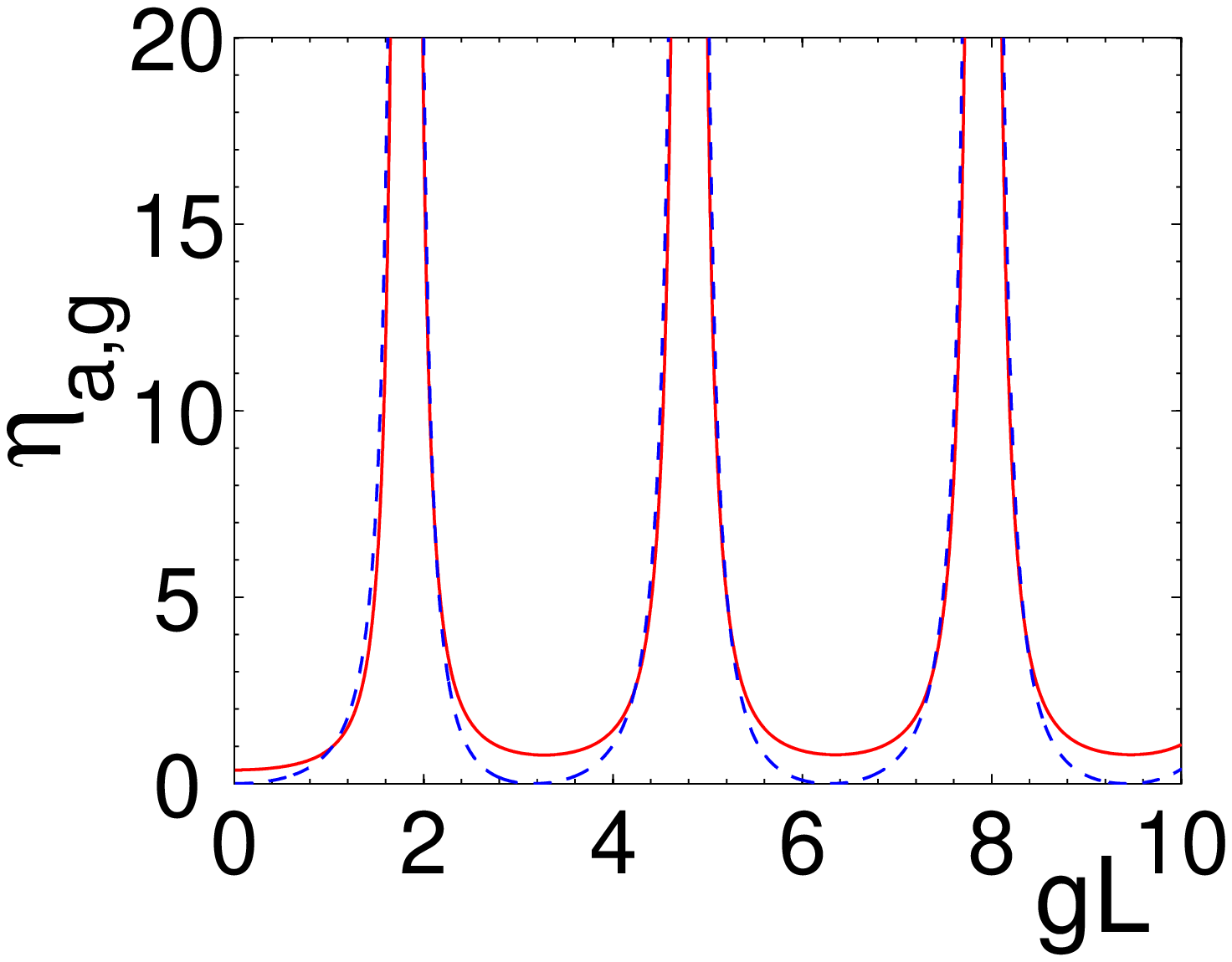} \includegraphics[width=.35%
\textwidth]{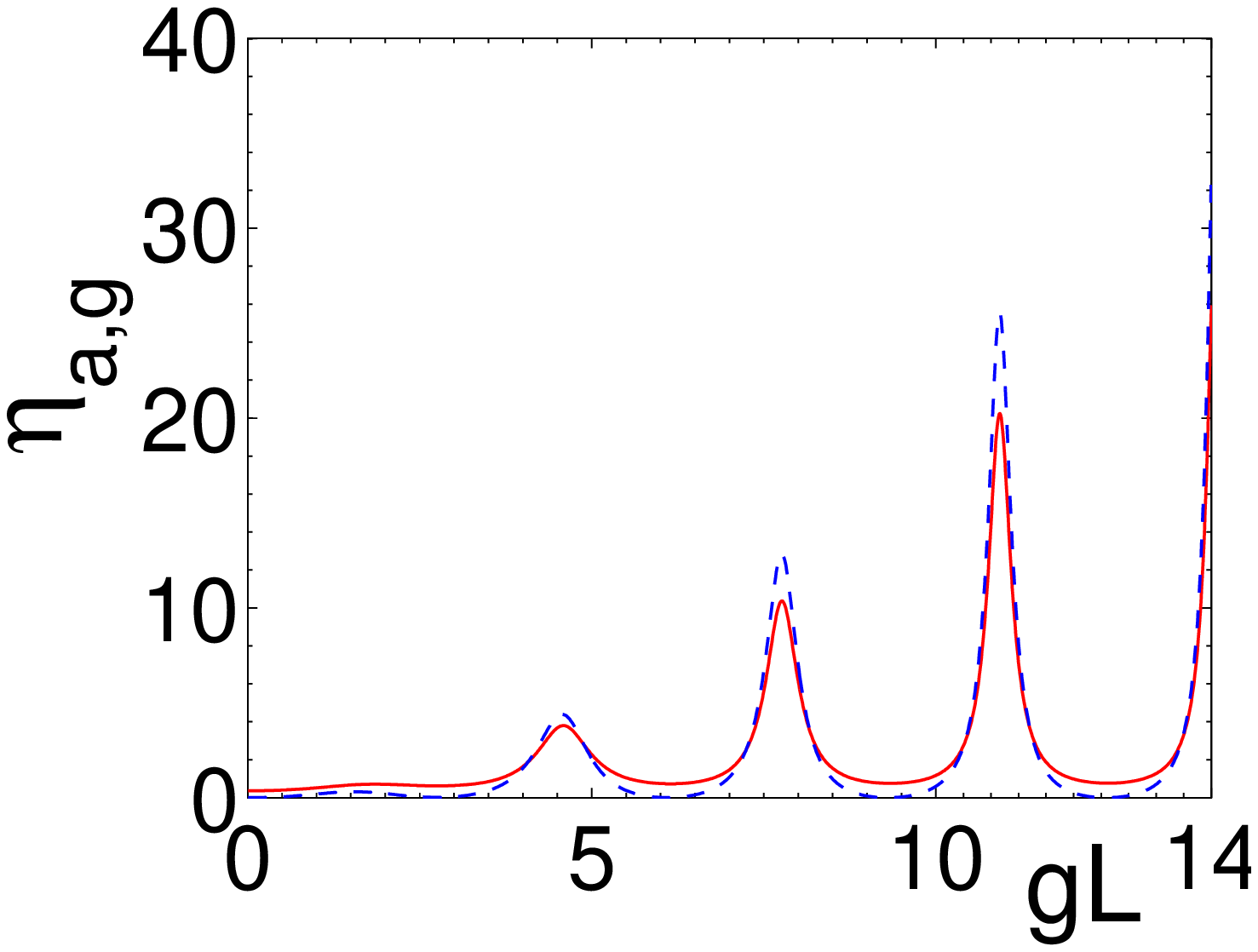} \caption{The amplification factor
$\protect\eta_{a}(\protect\omega_{1})$ (the solid line) and the
efficiency of difference-frequency generation $\protect\eta_{g}(
\protect\omega_{1})$ (the dashed line) for the backward wave at z=0.
$\protect\alpha_{1}L=1 $, $\protect\alpha_{2}L=1/2$. The upper plot:
$\Delta k=0$.\thinspace The lower plot $\Delta k=\protect\pi $.}
\label{fig2}
\end{figure}
\begin{figure}[h]
\includegraphics[width=.36\textwidth]{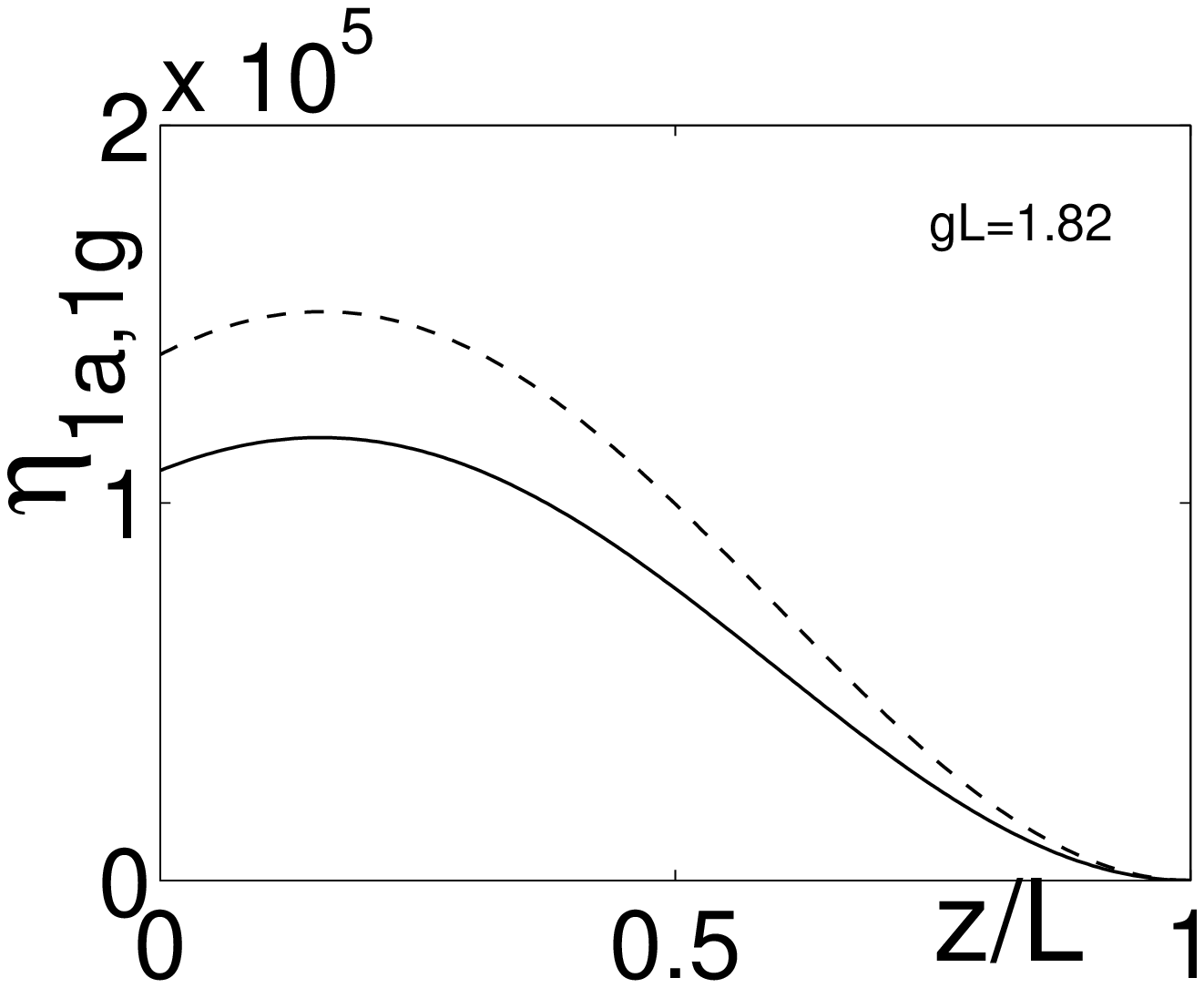} \includegraphics[width=.36%
\textwidth]{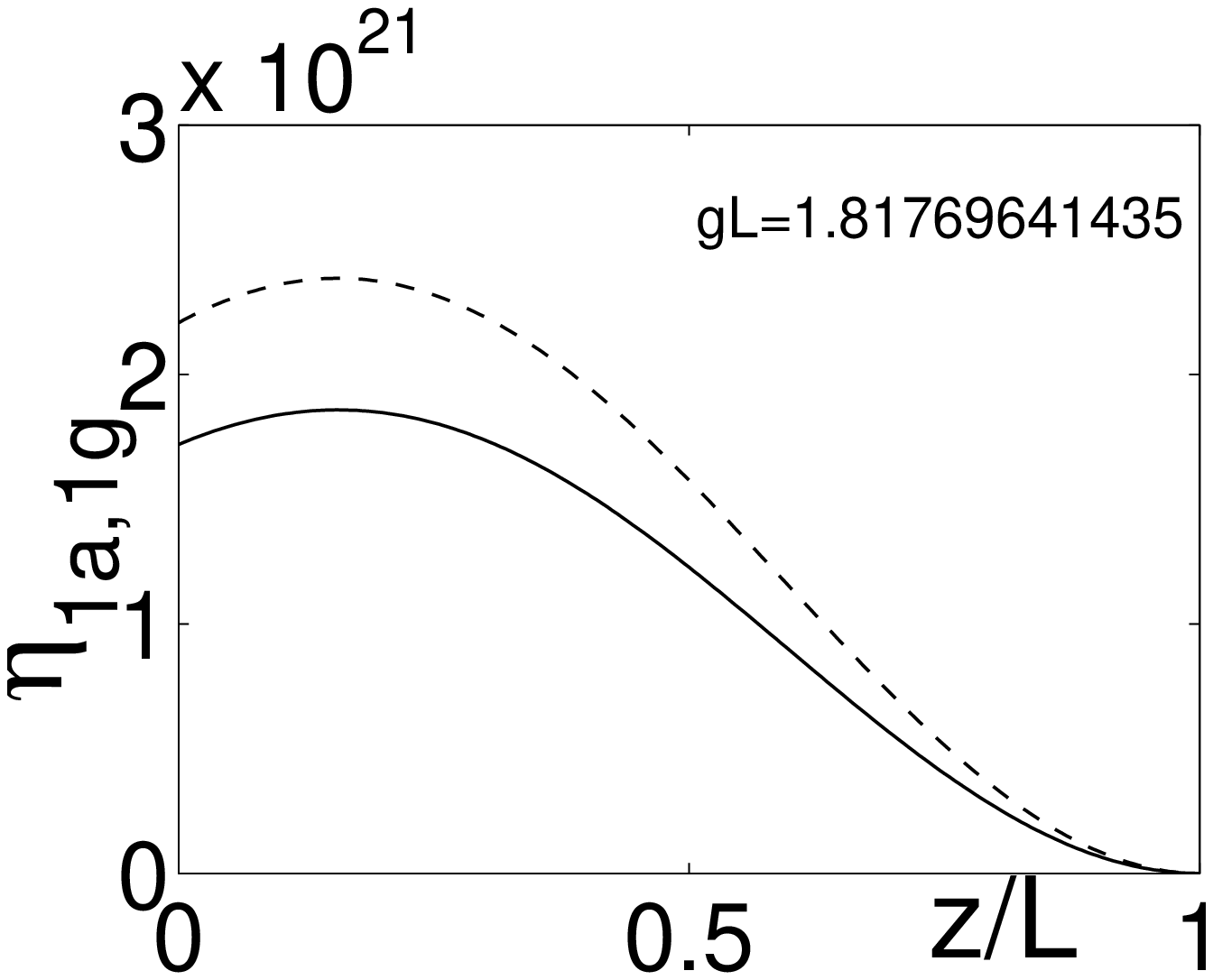} \caption{Resonant changes in the distribution
of the normalized intensity  of the left-handed wave inside the slab
of NIM, $\protect\eta_{a}(\protect\omega_{1})$ (the solid line) and
$\protect\eta_{g}( \protect\omega_{1})$ (the dashed line), caused by
the adjustment of the normalized intensity for the  control field at
$\omega_{3}$, $gL$. $\protect\alpha_{1}L=1$,
$\protect\alpha_{2}L=1/2$, $\Delta k=0$.} \label{fig3}
\end{figure}

The dependence of the output intensity for the left-handed wave
propagating in an absorptive NIM slab in the presence of the control
field at $\omega_{3}$ and at $a_{20}=0$ is shown with the solid line
in Fig. \ref{fig2} for two representative cases of exact and partial
phase-matching. The dash plots show the output in the case of DFG (at
$a_{1L}=0$, $a_{2,0}\neq 0$). Amplification in the upper part of Fig.
\ref{fig2} reaches many orders in the first maximum and increases in
the next maximums. It is seen that \emph{the amplification can
entirely compensate for absorption and even turn into oscillations}
when the intensity of the control field reaches values given by a
periodic set of increasing numbers. The larger the corresponding
value, the greater is the amplification and the DFG output; the latter
depends on the absorption for both waves and on the phase mismatch
$\Delta k$. The conversion factor is larger in its maximums than the
amplification factor because DFG is a one-step process, whereas OPA is
a two-step process as discussed in Subsection \ref{mrr}. As seen from
Fig. \ref{fig2}, the output shows a resonance dependence on the
intensity of the control field at $\omega_{3}$. Figure \ref{fig3}
depicts corresponding changes in the distribution of the
negative-index field inside the slab in the vicinity of the first
maximum at $\Delta k=0$.

\section{Conclusion}

We have studied the unusual properties of second-harmonic generation
(SHG) in metamaterials that have a negative refractive index for the
fundamental wave and a positive index for its second harmonic (SH).
The possibility of a left-handed nonlinear-optical mirror, which
converts the incoming radiation into a reflected beam at the doubled
frequency with efficiency that can approach 100\% for lossless and
phase-matched medium is considered. The most striking differences in
the nonlinear propagation and the spatial dependence of the
energy-conversion process for SHG in NIMs, as compared to PIMs, can
be summarized as follows. In NIMs, the intensities of the
fundamental and SH waves both decrease along the medium. Such
unusual dependence and the apparent contradiction with the ordinary
Manley-Rowe relations are explained by the fact that the energy
flows for the fundamental and SH waves are counter-directed, whereas
their wave-vectors are co-directed. Another interesting
characteristic of SHG in NIMs is that the energy conversion at any
point within a NIM slab depends on the total thickness of the slab.
This is because SHG in a NIM is determined by the boundary condition
for SH at the rear interface rather than the front interface of the
slab.

We have shown the feasibility of compensating losses in NIMs by
optical parametric amplification (OPA). In this process, the
wave-vectors of all three coupled waves are co-directed, whereas the
energy flow for the signal wave is counter-directed with respect to
those for the pump and the idler waves. This process is
characterized by properties that are in strict contrast with those
known for conventional nonlinear-optical crystals. Such
extraordinary features allow one to realize optical parametric
oscillations (OPOs) without a cavity at frequencies where the
refractive index is negative. We also showed that the OPA and OPO in
NIMs enable the generation of pairs of entangled counter-propagating
right- and left-handed photons inside the NIM slabs.

The backward energy flow for one of the coupled waves (whereas the
wave-vectors of all the coupled waves are co-directed) is inherent
for NIMs and it makes this process different from three-wave mixing
in PIMs. This is also different from various processes in RHMs based
on distributed gratings and feedback. The important advantage of the
backward OPA and OPO in NIMs investigated here is the distributed
feedback, which enables oscillations without a cavity. In NIMs, each
spatial point serves as a source for the generated wave in the
reflected direction, whereas the phase velocities of all the three
coupled waves are co-directed. As mentioned, it is very hard to
realize such a scheme in PIMs, while the OPA in NIMs proposed herein
is free from the PIM limitations.
\section{Acknowledgments}
The authors are grateful to V. V. Slabko for useful discussions and to
S. A. Myslivets for help with numerical simulations. This work was
supported in part by NSF-NIRT award ECS-0210445, by ARO grant
W911NF-04-1-0350, and by DARPA under grant No. MDA 972-03-1-0020.

\end{document}